\documentclass[twocolumn,epjc3]{svjour3}

\usepackage{graphicx,epsfig}
\usepackage{amsmath}
\usepackage{amsfonts}
\usepackage{bm}

\usepackage{braket}
\usepackage[hyperindex=true,colorlinks=true]{hyperref}
\usepackage[caption=false]{subfig}
\usepackage{multirow}
\usepackage{adjustbox}
\usepackage{scrextend}
\newcommand{\csubfloat}[2][]{%
  \makebox[0pt]{\subfloat[#1]{#2}}%
}
\newcommand{\centerhfill}[1][\quad]{\hspace{\stretch{1.}}#1\hspace{\stretch{1.}}}

\begin{document}

\title{Machine learning the deuteron: new architectures and uncertainty quantification}

\author{J. Rozalén Sarmiento\thanksref{ad:FQA,ad:ICC,em:JRS}
\and
J. W. T. Keeble\thanksref{ad:Surrey}
\and 
A. Rios\thanksref{ad:FQA,ad:ICC,ad:Surrey}
}

\thankstext{em:JRS}{\email{jrozalen@ub.edu}}

\institute{
\label{ad:FQA}
Departament de F\'isica Qu\`antica i Astrof\'isica, 
Universitat de Barcelona (UB), 
c. Mart\'i i Franqu\`es 1, E08028 Barcelona, Spain
\and
\label{ad:ICC}
Institut de Ci\`encies del Cosmos (ICCUB),
Universitat de Barcelona (UB), 
Barcelona, Spain
\and
\label{ad:Surrey}
Department of Physics, University of Surrey, Guildford GU2 7XH, United Kingdom
}

\date{Received: \today{} / Revised version: date}

\maketitle

\begin{abstract}
We solve the ground state of the deuteron using a variational neural network ansatz for the wavefunction in momentum
space. This ansatz provides a flexible representation of both the $S$ and the $D$ states, with relative errors in the
energy which are within fractions of a percent of a full diagonalization benchmark. 
We extend the previous work on this area in two directions. First, we study new architectures by adding more layers to the network and
by exploring different connections between the states.
Second, we provide a better estimate of the numerical uncertainty  by taking into account the final oscillations at the end of the minimization process. 
Overall, we find that the best performing architecture is the simple one-layer, state-disconnected network. 
Two-layer networks show indications of overfitting, in regions that are not probed
by the fixed momentum basis where calculations are performed. 
In all cases, the errors associated to the model oscillations around
the real minimum are larger than the stochastic initilization uncertainties.
\end{abstract}

\maketitle

\section{Introduction}
Artificial Neural Networks (ANNs) have become routine computational tools in a wide range of scientific domains~\cite{Mehta2019,Carleo2019}. 
Nuclear physics is no exception, and the number of machine learning (ML) tools is increasing steadily~\cite{NP_ML_2022,Utama2016,Gao2017,Regnier2019,Niu2019,Wang2019,Raghavan2021,Dong2021}.
An interesting and relatively recent use of ANNs is the solution of quantum mechanical problems, including in the many-body domain~\cite{Adams2021,Gnech2021}. 
These methods use ANNs as an ansatz for the many-body wavefunction,
and employ ready-made ML tools to minimise the energy of the system~\cite{netket2_2019,netket3_2021}.
The first attempts in condensed matter~\cite{Carleo2017,Saito2018} were quickly followed by advances both within~\cite{Vieijra2020} and outside the field~\cite{Gao2017}, including quantum chemistry ~\cite{Choo2020,Hermann2020}. A good example in this direction is FermiNet \cite{Pfau2020}, 
which uses an inherently
antisymmetric wavefunction as an ansatz in a variational approach to solve the many-electron problem. 

The methodology that we use to solve the deuteron here is based on the same variational philosophy
and stems from Ref.~\cite{Keeble2020}, 
where we used a one-layer variational ANN ansatz  
to describe
the wavefunction of the deuteron (the only $2-$body nuclear bound state) 
in momentum space. This yields excellent results in comparison with benchmark solutions 
for the energies and the wavefunctions, quantified through fidelity measures. 
Moreover, we estimate the out-of-sample uncertainty of the model by 
using several random initializations in the minimisation problem. 

The assessment of uncertainties in variational ANN (vANN) is particularly relevant to pinpoint the fundamental limitations of this methodology. 
There are several systematic errors that are important for vANN solutions to the Schr\"odinger equation. 
For instance, network architectures play a key role in determining the convergence properties of the network. 
Similarly, for a fixed architecture, vANN widths and depths change the expressivity 
of the network, and are thus fundamental ingredients in providing faithful
representations of continuous wavefunctions. 
One should explore this systematic uncertainties extensively before reaching conclusions
about any intrinsic shortcomings of the vANN method itself.

The deuteron is an excellent test bed for such studies. It is a relatively simple, one-body-like problem, which already includes the complexity of the strong force. Moreover, 
dealing with a wavefunction with two different angular momentum states adds an additional level of sophistication. 
In this manuscript, we extend the elementary approach
of  Ref.~\cite{Keeble2020} in two different directions. 
First, we look at somewhat more complex
ANN architectures to describe the wavefunction. This yields  insight on the importance of the network architecture and should ultimately provide information on the limitations of the method beyond a specific architecture. 
Second, we introduce a novel way to compute the out-of-sample error.
In particular, we look at oscillations in the energy, which is our cost function, around the real minimum. These oscillations
are an additional source of uncertainty, associated to the minimisation
process.
The analysis of such errors provides a new insight into uncertainty quantification for variational ANN solvers. 

This paper is structured in the following manner. In Section~\ref{sec:method} we briefly go over the methods used to approach the deuteron problem with ANNs. Section~\ref{sec:results} is devoted to the analysis of the numerical results, including their uncertainties.
Finally, in Section~\ref{sec:conclusions}, we draw a series of conclusions and comment on ways in which our work could be expanded and improved.

\section{Methodology}
\label{sec:method}
The approach that we use to solve the problem is variational. The ANN plays the role of a wavefunction ansatz, which we denote $\ket{ \psi_{\mathrm{ANN}}^{\mathcal{W}} }$.  
$\mathcal{W}=\{\mathbf{W^{(1)}}$,$\mathbf{W^{(2)}}$,$\mathbf{W^{(3)}}$,$\mathbf{b}\}$ 
is a set formed by network weights, $\mathbf{W^{(i)}}$, and biases $\mathbf{b}$. 
In a variational setting, the total energy of the system 
is the loss function, which reads, 
\begin{align}\label{eq:energy}
E^{\mathcal{W}}=\frac{\braket{\psi_{\mathrm{ANN}}^{\mathcal{W}}|\hat{H}|\psi_{\mathrm{ANN}}^{\mathcal{W}}}}{\braket{\psi_{\mathrm{ANN}}^{\mathcal{W}}|\psi_{\mathrm{ANN}}^{\mathcal{W}}}}.
\end{align}
The deuteron is the simplest nuclear two-body problem. 
We solve its structure in momentum space, where we can directly access nuclear interaction matrix elements.
We further separate 
the center of mass from the relative coordinates~\cite{Eisenberg1975}.  
Consequently, the wavefunction of the system depends only on the relative momentum coordinate, $\Vec{q}$. Working in momentum space allows us to skip numerically costly derivatives on the wavefunctions when computing the kinetic energy \cite{Raghavan2021}. 

We can further reduce the dimensionality of the deuteron problem via 
a partial wave expansion, thus separating the dependence on the absolute value of the 
momentum $q$ from its dependence on angles. 
For the deuteron, the tensor component of the strong interaction admixes the $S$ ($L=0$) and the $D$ ($L=2$) components of the 
ground-state wavefunction. 
The network will consequently have two different outputs, one for each state. 
The potential energy term used to compute Eq.~(\ref{eq:energy})  mixes these 
two states in a non-trivial way, so they are not completely independent from each
other in the variational setting. We choose the N3LO Entem-Machleidt nucleon-nucleon force \cite{entem-machleidt}, which is easily implemented in our code as a momentum-dependent potential. 

To implement the problem computationally, we use a one-dimensional grid in $q$ with $N_q=64$ points.
This grid is used to compute the energy integrals in Eq.~(\ref{eq:energy}) via Gaussian quadrature. 
The loss function is obtained from a global integral, the energy, and is always computed
with the aforementioned set of momentum grid points. As we shall see later, this can create some 
issues in deeper models. 
To efficiently capture the low-momentum structure and the high-momentum tails of the wavefunction, we use a mesh that densely
covers the region of low momenta but is sparse in regions of high momenta. We first distribute $N_q$ Gauss-Legendre points $x_i$ between $0$ and $1$, and then extend them tangentially using the transform 
$ q_i = \dfrac{q_{\mathrm{max}}}{\tan\dfrac{\pi}{2} x_N}\tan\dfrac{\pi}{2} x_i\,,$ with $i=1,\cdots,N_q$ and $q_{\mathrm{max}}=500$ fm$^{-1}$. In this set-up, the overlap $\braket{\psi_{\text{targ}}^L|\psi_{\text{ANN}}^L}$ 
, as well as all analogous integrated quantities, are discretised as follows:
\begin{align}
\braket{\psi_{\text{targ}}^L|\psi_{\text{ANN}}^L}
&= 4 \pi \int_0^\infty \!\!\! dq \, q^2 \, \psi_\text{targ}^{L *} (q) \psi_{\text{ANN}}^L(q) \nonumber \\
&\approx 4 \pi \sum_{i=1}^{N_q} w_i q_i^2 \psi_\text{targ}^{L *} (q_i) \psi_{\text{ANN}}^L(q_i) \,,
\label{overlap}
\end{align}
where $w_i$ are the integration weights associated to the tangentially-transformed Gauss-Legendre mesh. Note that
because the angular dependence has been removed via a partial wave expansion,
all the integrals involve a radial $q^2$ term.
Note also that all the wavefunctions are real-valued.  

In order to estimate the errors in our quadratures introduced by the discretization of the mesh, we have explored how different values of $N_q$ can affect values computed with this quadrature rule. For instance, computing the global energy, Eq.~(\ref{eq:energy}), with $N_q=48, 64$ and $80$ introduces variations of the order of keV. However, throughout the text we will use only $N_q=64$ and perform all comparisons with this number of points.

In the following, we present results corresponding to different ANN architectures for the two wavefunctions, $\psi_{\text{ANN}}^L(q)$, with $L=0$ 
and $2$.
Following Ref.~\cite{Keeble2020}, we take three identical steps to train all the networks. 
These three steps are common to most vANN problems in quantum mechanics~\cite{Saito2018}.
First, we initialise 
network parameters to uniform random values in the ranges 
$\mathbf{W^{(1)}}\in[-1,0)$, 
$\mathbf{W^{(2)}}\in[0,1)$ and 
$\mathbf{b}\in[-1,1)$. 
We choose a Sigmoid activation function, $\sigma(x)$, which provided marginally better results than the Softplus function in Ref.~\cite{Keeble2020}. We also explore the use of different activation functions $\sigma(x)$ in Sec.~\ref{subsection:further_exploration_of_hyperparameter_space}.

Second, we train the ANN to mimic a target function that bears a certain degree of physical meaning. 
We choose a functional form $\psi^L_{\mathrm{targ}}(q)=q^Le^{-\frac{\xi^2q^2}{2}}$, 
which has the correct low-momentum asymptotic values and a reasonable real-space width, $\xi=1.5\,\mathrm{fm}$. 
In this pre-training step, we use the overlap,
\begin{align}\label{eq:overlap}
    K^L = \dfrac{\braket{\psi_{\text{targ}}^L|\psi_{\text{ANN}}^L}^2}{\braket{\psi_{\text{targ}}^L|\psi_{\text{targ}}^L}\braket{\psi_{\text{ANN}}^L|\psi_{\text{ANN}}^L}},
\end{align}
to compute a total cost function, $C$, defined as,
\begin{equation}{\label{cost}}
    C=(K^S-1)^2+(K^D-1)^2\,.
\end{equation}
Equation~(\ref{cost}) is zero if, and only if, the two overlaps $K^S=K^D=1$. 
We choose the optimiser RMSprop~\cite{Mehta2019}, which dynamically adapts 
a global learning locally for all the network parameters. 
Training a medium-sized ANN with $N_{\mathrm{hid}}\approx 20$ hidden nodes 
usually takes about $10^4$ iterations. 
We point out that this pretraining step is not strictly necessary to achieve a correct energy minimisation, but it helps in guaranteeing a successful energy minimisation
with fewer  epochs. 

In the third and final step, we define a new cost function: the energy, as given in Eq.~(\ref{eq:energy}). We compute it with the momentum
quadrature described above. The network is then trained to minimise $E^{\mathcal{W}}$ for $2.5 \times 10^5$ epochs. The exact ground state energy can be computed in the very same momentum quadrature via exact diagonalization, and its value is $E_\mathrm{GS}=-2.2267$ MeV. We use this value as a benchmark.
We note that the kinetic energy has diagonal contributions 
in the $L=0$ and $2$ states, but the potential term mixes both states.
We perform all calculations in double precision. We use the PyTorch library~\cite{pytorch_paper,pytorch_docs}, which is particularly useful due to its automatic differentiation capabilities~\cite{PyTorch_AD}. A full version of the code is available on GitHub~\cite{Rozalen_github}.

\subsection{Architectures}

\begin{figure*}[t]
\centering
\hspace*{\fill}%
 \csubfloat[State-connected network with a single hidden layer. For this network, the number of parameters is $N=4N_{\mathrm{hid}}$, with $N_{\mathrm{hid}}$ denoting the number of hidden nodes. Note that the first layer includes a bias.]{\label{sc_1layer} %
  \includegraphics[width=0.45\linewidth]{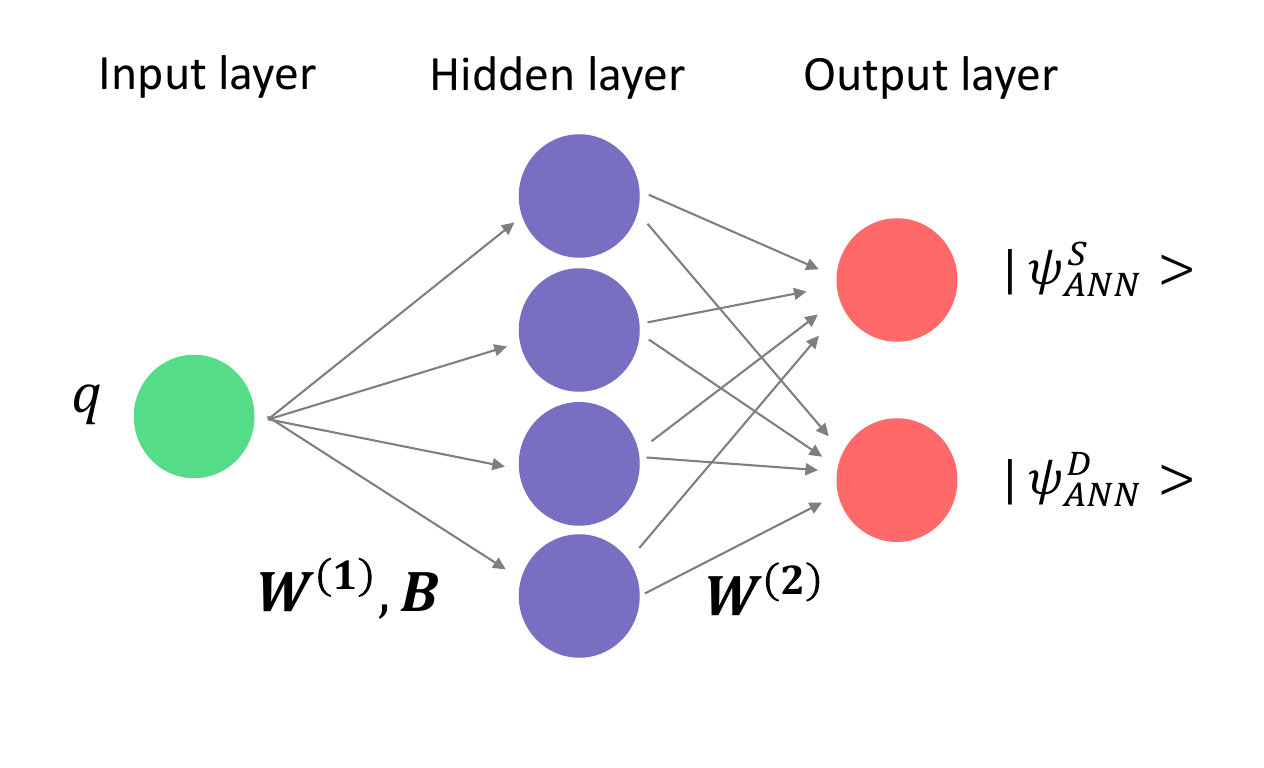}%
  }\centerhfill
  \csubfloat[State-connected network with two hidden layers, with $N=\frac{1}{8}N_{\text{hid}}(N_{\text{hid}}+12)$ parameters.]{\label{sc_2layers}%
  \includegraphics[width=0.45\linewidth]{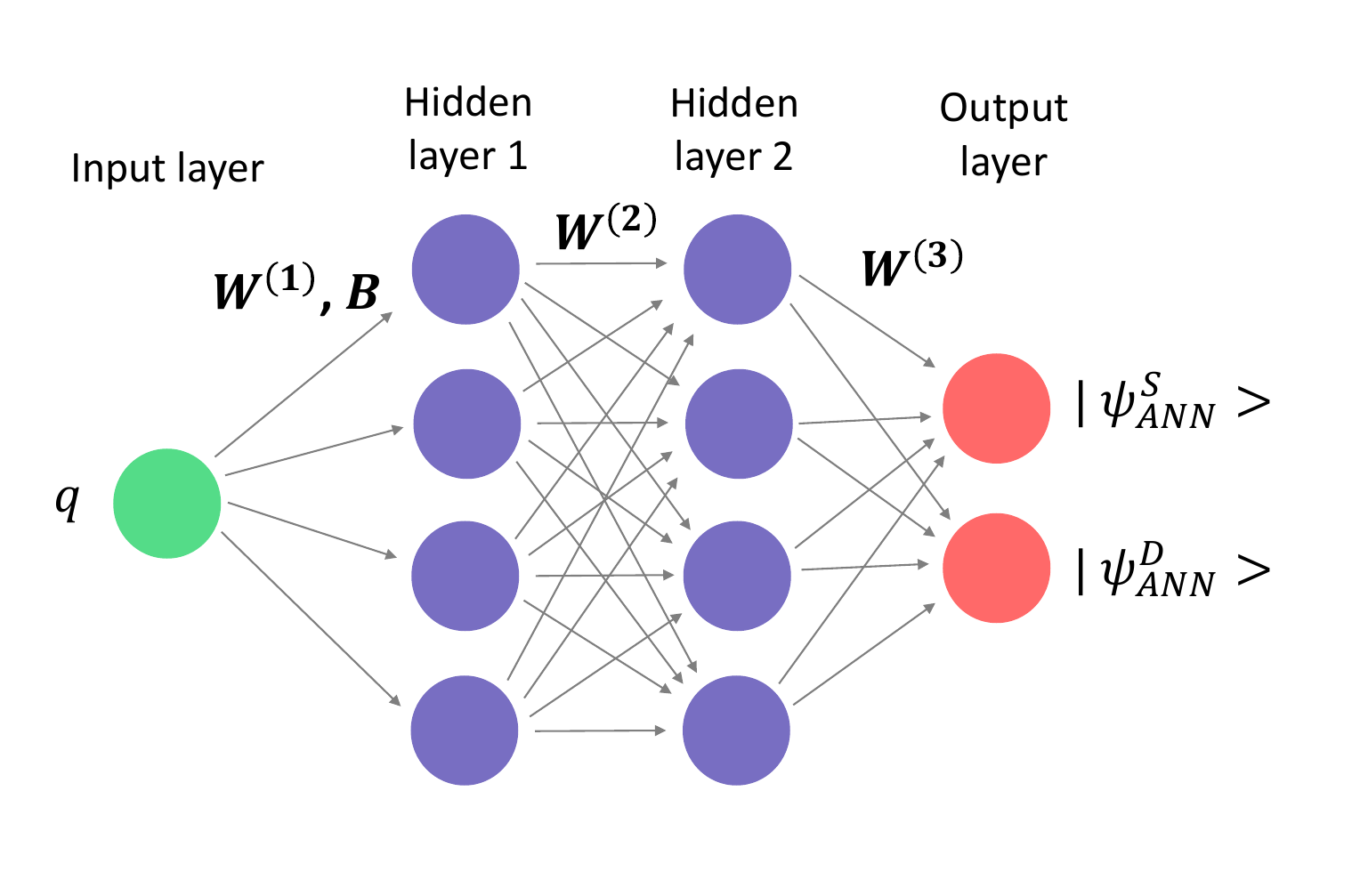}%
  }\hspace*{\fill}
  
  \hspace*{\fill}%
  \csubfloat[State-disconnected network with a single hidden layer, with $N=3N_{\mathrm{hid}}$ parameters.]{\label{sd_1layer}%
  \includegraphics[width=0.45\linewidth]{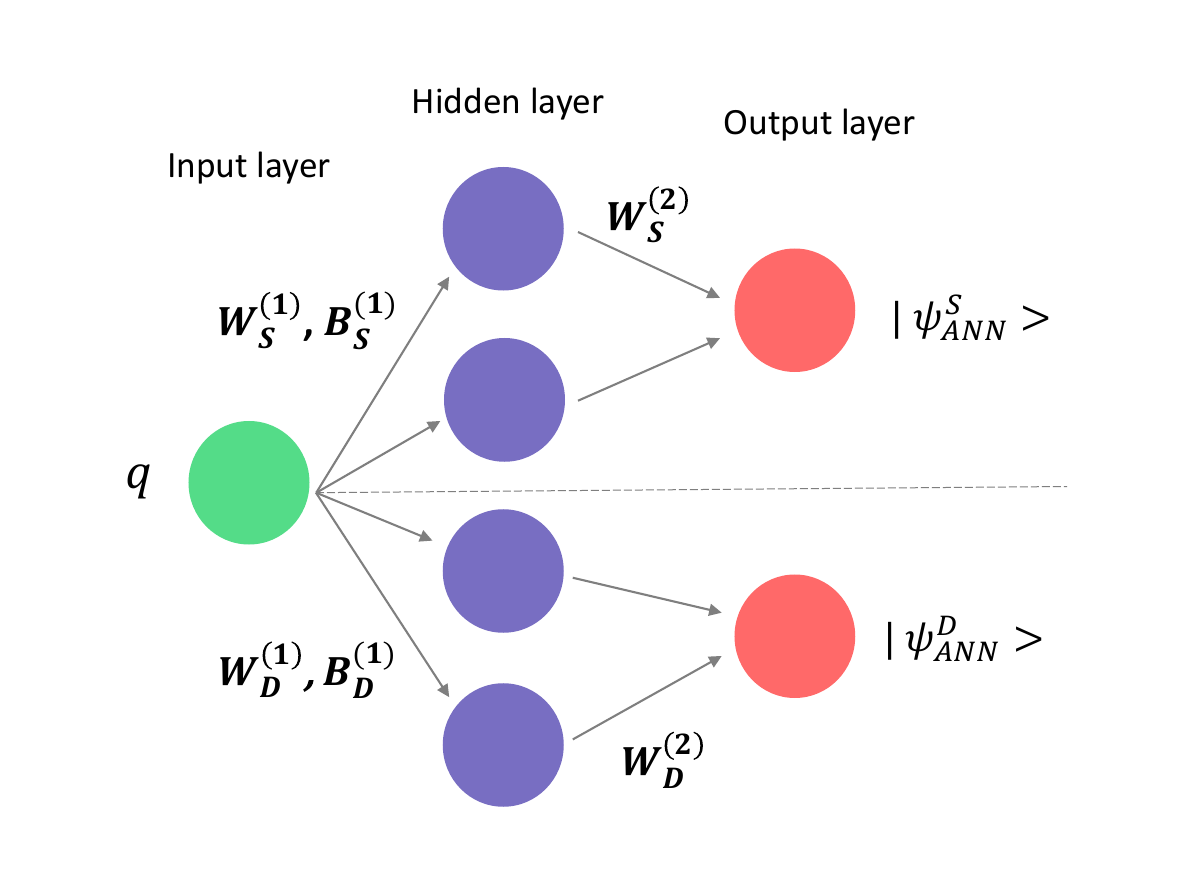}%
  }\centerhfill
  \csubfloat[State-disconnected network with two hidden layers, with $N=\frac{1}{4}N_{\mathrm{hid}}\left(N_{\mathrm{hid}}+8\right)$ parameters.]{\label{sd_2layers}%
  \includegraphics[width=0.45\linewidth]{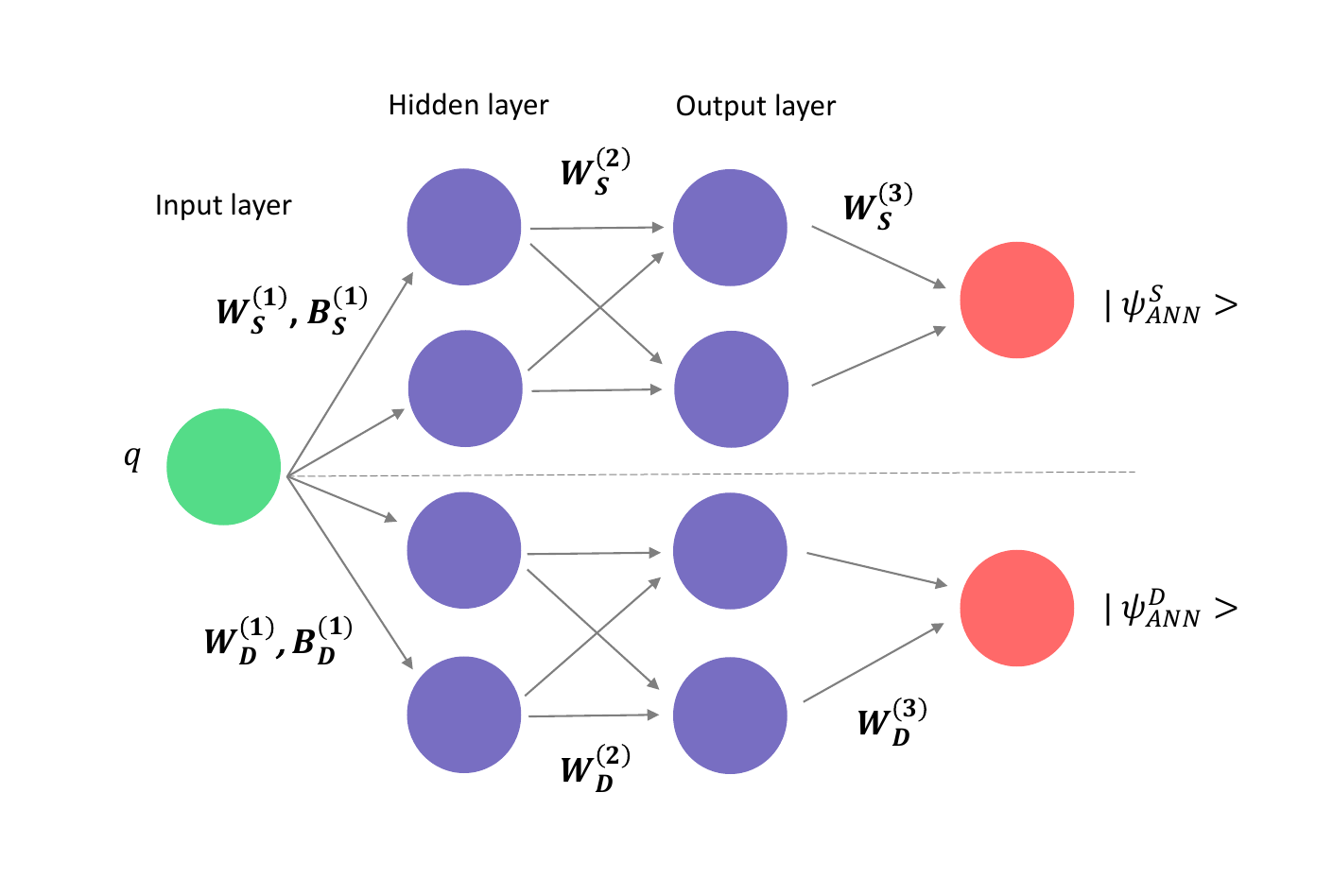}%
  }\hspace*{\fill}
  
  \caption{Neural network architectures used in this work.}
  \label{fig:archs}   
\end{figure*}

The vANN procedure is relatively general, and can be applied to any feed-forward ANN. In the initial exploration of deuteron 
wavefunctions of Ref.~\cite{Keeble2020}, a minimal approach was chosen deliberately to explore the potential and the limitations of the method in the simplest possible
model. 
We used a single-layer ANN with an increasing number of hidden nodes in order to assess systematics for a fixed architecture. 
Here, we take a step further and assess the impact of increasing the depth of the ANN 
which leads to different architectures.

To this end, we introduce four different ANN architectures, which we show in the 
four panels of Fig.~\ref{fig:archs}. 
All our networks have a single input, $q$, and two outputs, $\ket{ \Psi^S_\text{ANN} }$ and $\ket{ \Psi^D_\text{ANN} }$.
We distinguish between different configurations depending on how they connect to the two final states.
On the one hand, in fully-connected networks, which we dub ``state-connected'' (sc) networks, all the parameters contribute to both outputs
(see the top panels, Figs.~\ref{sc_1layer} and \ref{sc_2layers}). On the other hand, ``state-disconnected'' (sd)
 networks (bottom panels, Figs.~\ref{sd_1layer} and \ref{sd_2layers})
provide independent parameters for each state. One may naively expect sd networks to provide more flexibility (and hence
better variational properties) for the two wavefunctions.

In addition to the state dependence of the network, we also explore architecture systematics in terms of depth. We implement both sc and sd networks
with one (left panels) and two (right panels) layers. 
In the following, we use the acronyms $n$sc and $n$sd ($n=1,2$), with $n$ the number of layers, to refer to the 4 architectures explored in this work.
We stress that in Ref.~\cite{Keeble2020} we only looked at the $1$sc architecture. The motivation to explore other architectures and,
in particular further depths, is the widely-held belief that deep ANNs outperform their 
shallow single-layered counterparts~\cite{MacKay}. 

In the initial layer of all four architectures, we introduce both weights $\mathbf{W^{(1)}}$ and biases $\mathbf{b}$. In contrast, all the second (and/or third) layers only have weights, $\mathbf{W^{(2)}}$ ($\mathbf{W^{(3)}}$). 
For clarity and completeness, we now provide the functional form of the four networks. For the one-layer $1$sd configurations, 
the wavefunction ansatz reads,
\begin{align}{\label{ann1_equation}}
    \psi^L_{\mathrm{ANN}}(q)=\sum^{N_{\mathrm{hid}}/2}_{i=1}W^{(2)}_{i,L}\sigma\left(W^{(1)}_{i,L}q+B_{i,L}\right).
\end{align}
The corresponding $1$sc contribution is obtained by setting $N_{\mathrm{hid}}/2 \rightarrow N_{\mathrm{hid}}$ and 
using $L-$independent $\mathbf{W^{(1)}}$ and $\mathbf{B}$  parameters.
When $2$ layers are included, in contrast, the sd ansatz is more complex, and becomes the sum of sums of two nested activation functions,
\begin{align}{\label{ann2_equation}}
    &\psi_{\mathrm{ANN}}^L(q)= \nonumber \\
    &\sum_{i=N_{\mathrm{hid}}/2+1}^{N_{\mathrm{hid}}}W^{(3)}_{i,L}\sigma\left(\sum_{j=1}^{N_{\mathrm{hid}}/2}W^{(2)}_{j,L}\sigma\left(W_{j,L}^{(1)}q+B_{j,L}\right)\right).
\end{align}

One can again easily manipulate the previous expression to obtain a $2$sd wavefunction.

The relation between the total number of hidden neurons, $N_{\mathrm{hid}}$, and the total number of variational parameters, $N$, is different for each architecture. 
The relations $N(N_{\mathrm{hid}})$ are shown in the captions of Fig.~\ref{fig:archs} for each network configuration. 
One-layer networks have a linear $N(N_{\mathrm{hid}})$ relation, whereas the relation for two-layer networks is quadratic. In other words,
for the same $N_\text{hid}$, two-layer networks will usually involve a much larger 
number of parameters than one-layer models, and one may expect overfitting to
become an issue. Whereas for the $1$sc architecture the number of hidden nodes is unrestricted so long as $N_\text{hid}>2$, the $2$sc 
architecture must have $N_\text{hid}>4$ with $N_\text{hid}$ even. Likewise, the $1$sd network must have an even $N_\text{hid}$.
In contrast, for the $2$sd network, $N_\text{hid}$ is a multiple of $4$.

\subsection{Learning process}
We minimise the energy cost function using RMSprop~\cite{Mehta2019} with hyperparameters $\epsilon=10^{-8}$ and a smoothing constant $\alpha=0.9$.
We set the learning rate to $10^{-2}$ and explore the $N_{\mathrm{hid}}$ dependence using the values
$
    N_{\mathrm{hid}}\in\{20,30,40,60,80,100\}
$ (in the 2sc architecture we change $N_{\mathrm{hid}}=30\rightarrow N_{\mathrm{hid}}=32$).
To explore the full flexibility of the wavefunction ans\"atze, we optimise the networks 
by mimising the overlap loss function for
$2 \times 10^3$ epochs of pre-training, followed by 
$2.5 \times 10^5$ epochs of energy minimisation.
Rather than doing this a single time, we use $20$ different random initializations and display mean values
and (Bessel-corrected) standard deviations obtained with all these runs\footnote{This is to be compared to the $50$ runs shown for
the $1$sc  architecture in Ref.~\cite{Keeble2020}.}. 
With this, we explore the out-of-sample bias of the network
and we attempt to draw generic conclusions for the network architecture, rather than for a single, specific network model.

\begin{figure}[t]
    \includegraphics[width=\linewidth]{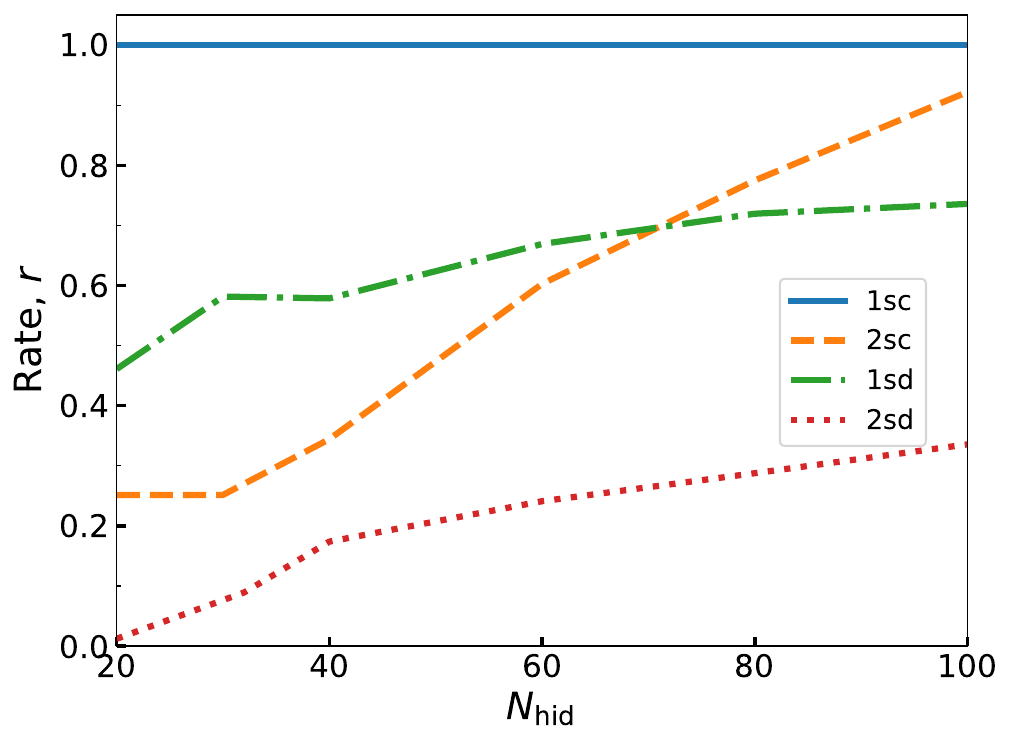}
    \caption{Convergence rate of the four network architectures as a function of the number of hidden neurons. The rate (vertical axis) is defined as the ratio between the number of models that converge and the total number of trained models, $N_\text{con}/N_\text{tot}$.}
    \label{fig:convergence_rate}
\end{figure}

In fact, we perform $150$ minimisations for each architecture (other than for a specific $2$sd model, as we discuss in the following). 
Not all of these runs converge or, if they
do, they may not converge to meaningful values, close enough to the minimum. 
Figure~\ref{fig:convergence_rate} shows the rate of convergence of different architectures 
as a function of $N_\text{hid}$. 
This is obtained as the
ratio of the number of converged models, $N_\text{con}$, to the total set of model 
initializations, $N_\text{tot}$=150, $r=N_\text{con}/N_\text{tot}$. 
As a selection criterion to compute $N_\text{con}$, 
we define converged models as
those that provide a final energy within the range  $E\in (-2.220,-2.227)$ MeV. 
We set $N_\text{tot}=150$ to guarantee that all
configurations have a minimum of $N_\text{con}=20$ converged models. 
The only exception is the $2$sd architecture with $N_\text{hid}=20$, which has a very small
convergence rate. This architercture requires $N_\text{tot} = 1100$ runs to get 
$N_\text{con}=20$ converged models.
Whenever more than $20$ initializations lead to converged results, we randomly pick $20$ states to have representative, but also
homogeneous, statistics across the $N_{\mathrm{hid}}$ domain. 

We stress the fact that these rates are obtained from initializations with a pretraining step. With no pretraining, all models 
show slower convergence rates\footnote{This includes the $1$sc configuration, 
which leads to a perfect convergence rate if pretraining is included.}. 
We take this difference as an indication of the fact that pretraining is effective in providing a physical 
representation of the ANN wavefunction. 
Initally non-pretrained networks may occasionally lead to accurate results, but are likely
to require many more energy minimisation epochs. 

Going back to Fig.~\ref{fig:convergence_rate}, 
we find that the one-layer configurations provide better convergence rates than their two-layer 
counterparts up to 
$N_{\mathrm{hid}} \approx 60$, at which point
the $2$sc architecture is marginally better than $1$sd. The $1$sc network (solid line) 
has a perfect convergence rate. The $1$sd architecture (dash-dotted line) 
provides a convergence rate of $r > 50 \%$ relatively independent of the value of $N_\text{hid}$. We attribute these similarities 
to the fact that both pretraining and training are more 
easily done with a single-layer ANN. 

In contrast, two-layer networks start with a relatively low 
convergence rate. In architecture $2$sc we have $r \approx 25 \%$ for $N_\text{hid}=20$, and the rate increases in a relatively steady and linear fashion up to 
$r \approx 80-90 \%$ for $N_\text{hid}=100$. In architecture $2$sd we find even lower convergence rates, $r \approx 1 \%$ for $N_\text{hid}=20$, and a relatively slow increasing tendency in terms of $N_\text{hid}$ up to $r \approx 20 \%$. 
The stark differences in convergence rates between one and two-layer 
ANNs may be due to a variety of factors. 
On the one hand, network parameter initialization may be an issue. 
Network parameters in different domains than the ones we prescribe at the moment may provide better starting points. 
On the other hand, we fix the total number of energy minimisation 
epochs and the current rate may just be an indication that smaller two-layer
networks simply take, with RMSprop, a larger number of epochs to converge.
Finally, two-layer networks may be problematic in terms of the Sigmoid activation function, which is prone to a vanishing gradients issue~\cite{Hochreiter1998,Bengio1994}. This
should be accentuated in deeper (as opposed to shallow one-layer) networks. 
Along these lines, having few neurons increases the probability that all of them become 
frozen, which may explain why low $N_\text{hid}$ models have a harder time converging. 
These arguments suggest that an activation function like Softplus or ReLU work better with $2$-layer architectures. We have carried out tests with different activation functions that show an improvement in convergence rates for $2-$layer networks, as shown in Table~\ref{tab:hyperparams2}.

\subsection{Error analysis}
In addition to different network architectures, we examine two 
types of uncertainties. First, as we have just explained, we
run $20$ minimizations for different $N_\text{hid}$ values and 
network architectures. We take the standard deviation of these $20$ results as a measurement
of the uncertainty related to the stochastic minimization process. This 
out-of-sample error, which was also explored in Ref.~\cite{Keeble2020}, 
is represented in terms of dark bands in the figures of the following section. It tends to be a relatively small error, of the order of a fraction of a keV in energy. 

Second,  even after a full minimisation,  our results still show 
some residual oscillations. These oscillations are typically within 
a few keV of the total energy. 
Rather than keeping only the lowest values of the energy as our final values, we instead adopt a more conservative stance and we average over the oscillations to obtain a central value. 
The oscillation amplitude can then be used to quantify an 
additional source of uncertainty, associated to the minimiser. 
To determine this error, we take a trained model and let it evolve for $300$ additional epochs - a process which 
we call post-evolution. The number of post-evolution epochs is
small enough so that the 
mean value does not change during the process, but also large 
enough to observe periodicity in the oscillations.

\begin{figure}[t]
    \includegraphics[height=0.29\textheight]{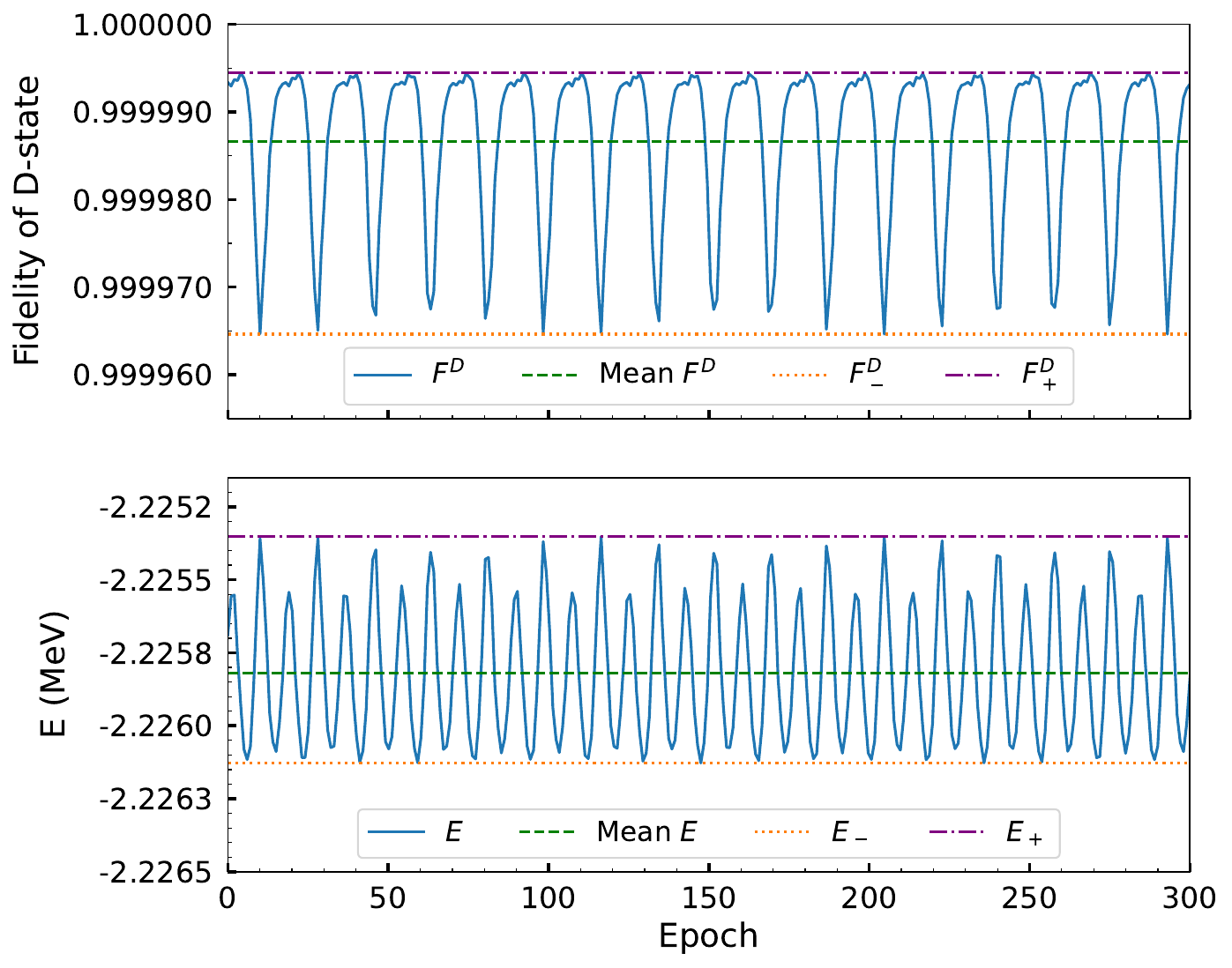} 
    \caption{Top panel: D-state fidelity post-evolution of a trained model with architecture $1$sc and $N_{\mathrm{hid}}=100$ hidden neurons. 
    Bottom panel: total energy evolution of the same model as in the top panel.}
    \label{fig:oscillation_error}
\end{figure}

We illustrate this behavior in Fig.~\ref{fig:oscillation_error}. The top panel shows the evolution of the $D-$state fidelity
$F^D$ over the $300$ post-evolution epochs for a $1$sc network with $N_\text{hid}=100$. Here, we define $F^L$ as the overlap between the ANN and benchmark (obtained via exact diagonalization) wavefunctions in analogy to Eq.~(\ref{eq:overlap}).
The  solid blue line, corresponding to the fidelity at every epoch, displays clear oscillations that are asymmetric
with respect to a mean value of $F^D \approx 0.999987$. To account for the asymmetry in the oscillating behavior,
we assign an upper and lower error estimate. 
The upper value corresponds to the maximum value across
the post-evolution phase - $\delta F^D_+=0.000008$ (dash-dotted line), in the case shown
in the figure. In contrast, the lower value is significantly lower for this
specific example,
leading to $\delta F^D_-=0.00002$ (dotted line). 
The bottom panel of Fig. \ref{fig:oscillation_error} shows the energy (solid line) 
as a function of the post-evolution epoch for the same simulation. The energy oscillates around the mean value 
(dashed line), and the top and bottom bounds of these oscillations are shown in 
dash-dotted and dotted lines respectively. In this particular example, the mean value 
of the energy is $E^{\mathcal{W}}\approx -2.2258\,\mathrm{MeV}$, with oscillation errors 
$\delta E^{\mathcal{W}}_+=0.5\,\mathrm{keV}$ and 
$\delta E^{\mathcal{W}}_-=0.3\,\mathrm{keV}$. We note that these values are typical  for almost any number of neurons, and are extremely small, less than 
$0.1 \%$ of the total energy value.

In order to take into account the stochastic out-of-sample uncertainty
in these oscillations,
we repeat the process above 
several times for each network configuration 
($N_{\mathrm{hid}}$) and keep $20$ random samples of the networks. 
As explained earlier,
we extract central values for the different physical quantities by taking the average 
over these $20$ mean values. We quote post-evolution 
oscillation errors that also include this stochastic element. In other words, 
both upper and lower bounds are calculated by averaging over 20 individual 
post-evolution runs. In a cost function with no near multiple minimums of similar 
heights, all oscillations should happen around the same minimum, and the mean values of 
$E^{\mathcal{W}}$ should all be similar. This, in tandem with the fact that we estimate 
the stochastic $\delta E^{\mathcal{W}}$ as the standard deviation, leads us to expect 
small stochastic errors and comparatively bigger oscillation errors.
We confirm this expectation in the following section. 

\section{Results}
\label{sec:results}
\subsection{Energy}

\begin{figure*}[t]
        \includegraphics[width=\linewidth]{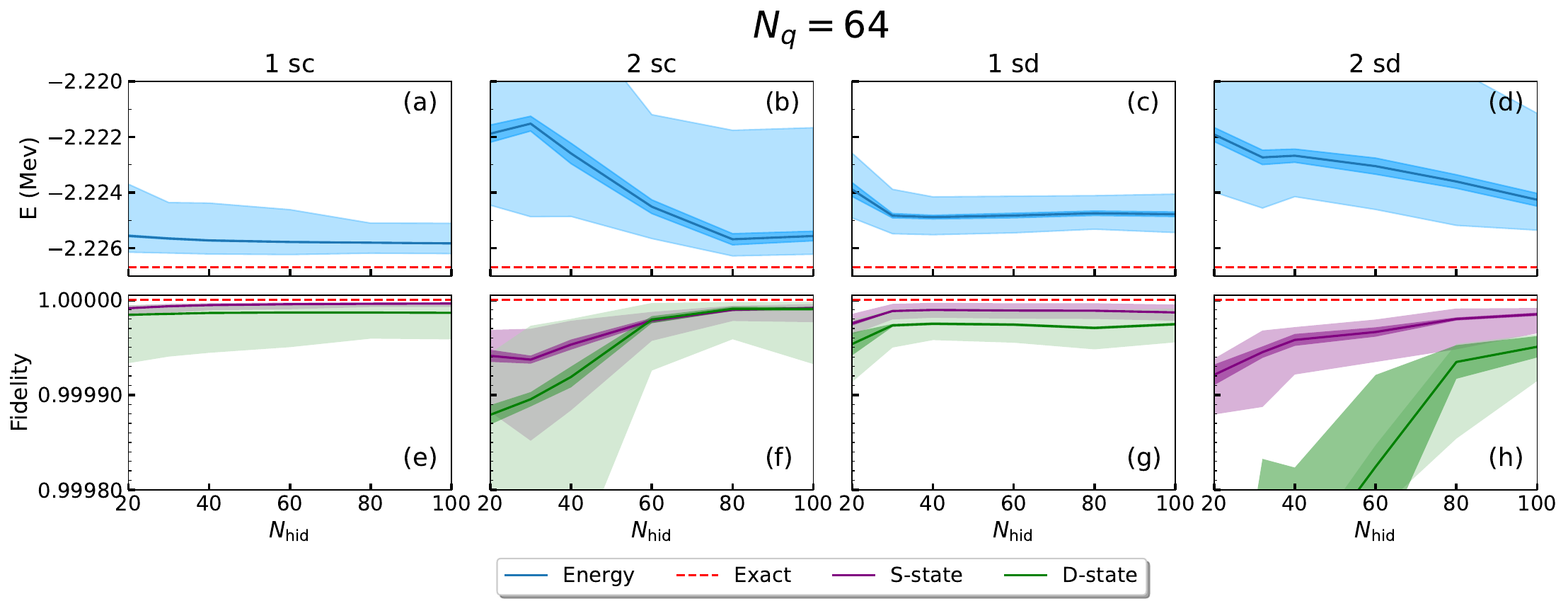}
        \caption{
        Top panels: energy as a function of the number of hidden nodes. Different panels
        correspond to the network architectures of Fig.~\ref{fig:archs}. Lines represent
        central values obtained with $20$ stochastic initializations. Dark bands show the associated standard deviations (stochastic uncertainty), whereas light bands show the post-evolution oscillation uncertainty. The dashed red lines indicate the benchmark value.
        Bottom panels: the same for the fidelity of the ANN wavefunction with respect to the benchmark for the $S$ (purple) and $D$ (green) states.
        }
        \label{fig:energy_fidelity}
\end{figure*}

The two quantities that we use as indicators of the quality of our final trained models are the energy and the fidelity, represented in 
the top and bottom panels of Fig.~\ref{fig:energy_fidelity}, respectively. 
The four columns correspond to the different network architectures of Fig.~\ref{fig:archs}. 
The solid dark lines in Fig.~\ref{fig:energy_fidelity} indicate the central values. 
We show two different types of uncertainties for each quantity. 
First, stochastic uncertainties associated to 
$20$ different initializations are shown using dark colour bands.
Second, post-evolution oscillation uncertainties are displayed with pale colours.
A key finding of our work is that post-evolution uncertainties are always larger than stochastic uncertainties. 

Before discussing uncertainties, we want to stress the quality of the results. As discussed in the context of Fig.~\ref{fig:convergence_rate}, the ANNs used to generate these results have been preselected to lie
in the range  $E\in (-2.220,-2.227)$ MeV. Therefore, all these 
models already lie within $0.3 \%$ of the final energy
result. For some architectures, the rate of convergence to this energy
window is not $100 \%$. Yet, converged results provide energy values which,
according to the stochastic uncertainty, are well within $1$ keV (or $0.04 \%$) of each other. 
Post-evolution uncertainties are much larger than stochastic uncertainties, but are 
typically smaller than $4$ keV (or $0.2 \%$). 
We take this as an indication that all these ANNs provide faithful, high-quality 
representations of the deuteron wavefunction. 
While none of the models shown here are able to reach the real minimum even when the 
post-evolution lower bounds are considered, the distance between the lower bound
and the real minimum is always less than $2$ keV. 
Our analysis indicates that this limitation is due to the fact
that the network cannot always distinguish between the
$S-$ and $D-$state contributions in high-momentum regions, where the
two wavefunctions are equally small. Ultimately, this limitation only has 
minor consequences,
of a fraction of a percent, in the total energy. An important takeaway of this 
analysis is that the simplest architecture, $1$sc, provides the most stable
and overall best performing results: see panel (a) of  Fig.~\ref{fig:energy_fidelity}.

An interesting conclusion of Fig.~\ref{fig:energy_fidelity} is that state-disconnected architectures (panels (c) and (d)) perform seemingly worse than state-connected networks. One could have expected 
the opposite, on the basis that disconnected architectures may 
provide more independent flexibility for the $\psi^S$ and $\psi^D$ wavefunctions. 
One possible explanation for this behavior is as follows. Within a single minimisation 
epoch, the optimiser proposes steps based on the backpropagation of gradients. Changes 
in parameters associated to the $S-$state are thus immediately propagated 
to the $D$ state in a fully connected configuration. In contrast, in the disconnected 
case, these changes do not necessarily affect 
the $D$ state. In spite of having fewer parameters than connected 
architectures, the training may take longer due to this effect. In a minimisation protocol with a fixed number of epochs like ours, this may lead to worse results. 

We now proceed to discuss the $N_\text{hid}$ dependence of the results. We find two very different trends depending on the depth of the network. One-layer network results are 
relatively independent on the network width.
In other words, in terms of energy (but also in terms of fidelity), models with $N_\text{hid} \approx 30$ are as good as models with $N_\text{hid} = 100$, as can be seen in panels (a) and (b) of Fig.~\ref{fig:energy_fidelity}.
In contrast, the two-layer models display a clear improvement in energy. This suggests that two-layer networks with small values of $N_{\mathrm{hid}}$ have
not managed to reach the full minimum after $2.5 \times 10^5$ epochs. 
Single-layer models have notably fewer parameters and can thereby be trained faster than their two-layer counterparts. 
In fact, snapshots of the minimisation process indicate that two-layer models with few neurons are still learning at 
the end of the $2.5 \times 10^5$ epochs. 
The decrease in energy as $N_{\mathrm{hid}}$ increases extends to the whole range of $N_{\mathrm{hid}}$ in architectures $2$sc
and $2$sd (panels (b) and (d) of Fig.\ref{fig:energy_fidelity}).
However, the post-evolution uncertainties for these two-layer models
are relatively big and almost compatible with a constant 
value of energy. 

In addition to central values, the dependence of uncertainties in $N_\text{hid}$ is 
informative. The stochastic uncertainties
are relatively constant across all values, but are marginally larger for two-layer 
models. Post-evolution errors for 
one-layer networks are on the order of $\approx 2$ keV, with a mild decreasing 
dependence on network width. 
In contrast, two-layer network post-evolution uncertainties are as large as $6$ ($4$) 
keV for low (high) $N_\text{hid}$ values. 
Following the same argument about the energies of such models, this can be understood assuming that the updates in $\mathcal{W}$ from RMSprop affect both $\psi^S$ and $\psi^D$ in the state-connected case (as opposed to the state-disconnected case).
One expects this may lead to larger energy variations and, hence, 
bigger oscillations. 

\begin{figure*}[t]
    \includegraphics[width=\linewidth]{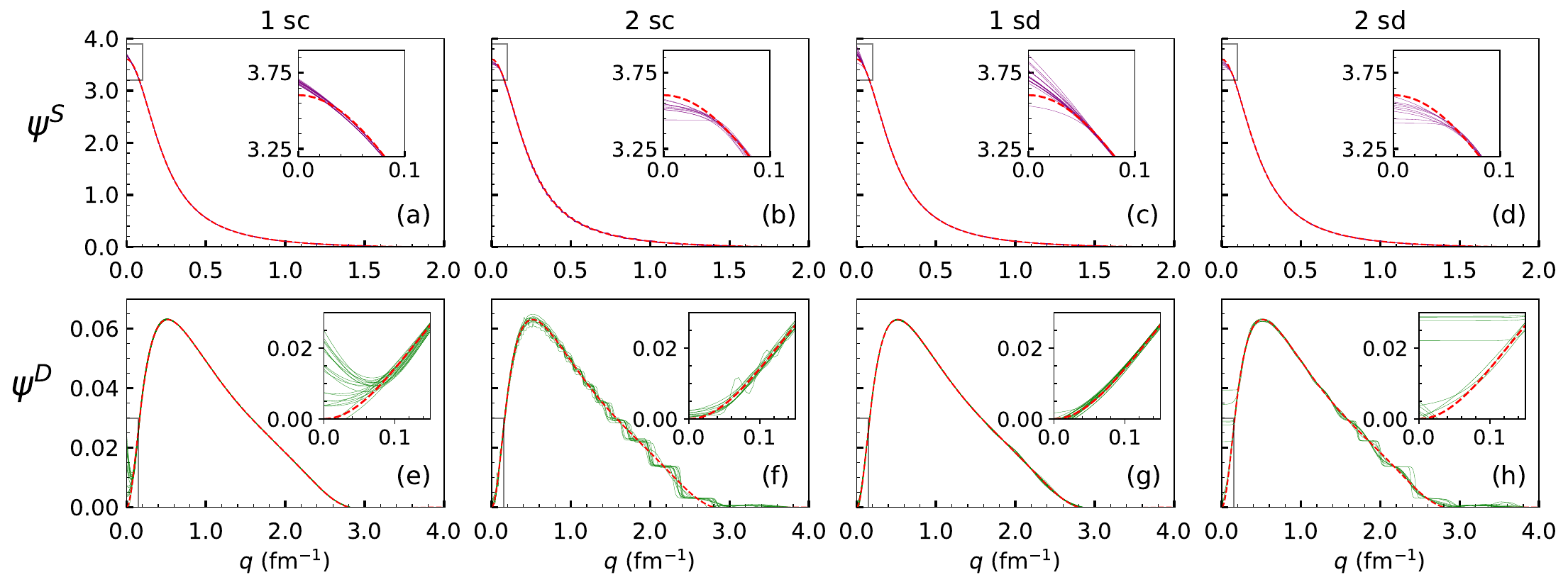}
    \caption{
    Top panels: $20$ instances of the $S-$state wavefunction as a function of momentum, for each of the 4 architectures considered here for $N_\text{hid}=100$.
    The dashed line represents the benchmark wavefunction. 
    The insets focus on the region around the origin, $q \approx 0$. 
    Bottom panels: the same for the $D-$state wavefunction. Note the difference in scales of the momentum in both rows.
    }
    \label{fig:wf_var_wf}
\end{figure*}

\subsection{Fidelity}
The panels (e)-(h) at the bottom of Fig.~\ref{fig:energy_fidelity} also provide an insight on the model quality after minimisation. 
Again, we highlight that overall the fidelities are extremely close to benchmark values, within a fraction 
of a percent in all cases. 
We find general conclusions in terms of network width that 
are in line with those we found for the energy. 
Across all architectures and network depths, the fidelity of the $S-$state is better than that associated to the
$D$-state. Not only are the central values of $F^S$ closer to one, but post-evolution 
uncertainties are significantly
smaller for this state. This is in contrast to the much smaller stochastic uncertainty, 
which is of the same
size for both states\footnote{The only exception is the $D-$state fidelity of 
the $2$sd architecture with $N_\text{hid}=60$, shown in panel (h) of Fig.~\ref{fig:energy_fidelity},
which has a larger oscillation uncertainty.}. 

When it comes to the $N_\text{hid}$ dependence of the results, we find again that 
one-layer networks are relatively independent
of the network width. Overall, networks perform better 
with $1$sc than with $2$sc architectures
as measured by overlaps that are closer to one, although they have 
relatively similar uncertainties. The 
fidelities of the $1$sd architecture are relatively constant as the width 
increases. The behavior of the $2$sd networks shown in panel (h) is more erratic. While the $S-$state 
fidelity is relatively close
to the benchmark and seemingly improves with $N_\text{hid}$, the $D-$state 
fidelity is closer to one for 
$N_\text{hid}=60$, and subsequently its quality decreases. We take this as a sign that 
such architectures with 
$N_\text{hid}$ values larger than $80$ may be problematic and, in fact, we show in the 
following subsection that overfitting
is an issue in this region. Finally, we stress again that there is no observable 
bias-variance trade-off in the fidelities \cite{Mehta2019}. 

\subsection{Wavefunctions}

We now turn our attention to analyzing directly the ANN outputs: the wavefunctions. This is an instructive exercise that
allows us to have local indicators of model quality, in addition to the information provided by integrated
quantities such as the energy or the fidelity. 
We show $20$ different instances of the wavefunctions of the $S-$ (top panels, (a)-(d)) and $D-$states 
(bottom panels, (e)-(h)) in Fig.~\ref{fig:wf_var_wf}.     
To show the ``best" possible wavefunctions, we focus on the number of hidden nodes $N_\text{hid}$  that provide the
lowest (central value of the) energies. 
For all four architectures considered, this corresponds to $N_\text{hid}=100$.
By looking at these 
instances directly, one gets an idea of the possible spread of 
variationally evolved models. 
Overall, we find that all networks provide very similar, high-quality
representations of the benchmark wavefunction, which is shown in dashed lines.

In most cases, the largest discrepancies occur towards the origin. As explained in 
Ref.~\cite{Keeble2020}, the presence
of a $q^2$ factor in the energy integrals allows the networks to push a large amount of 
variance towards the low-$q$ region. In other words, changing the wavefunction near
the origin has no reflection 
on the global cost function. 
The large variance can be seen in the insets of 
all panels. For the $S-$state, one-layer models have relatively linear behaviors as $q \to 0$,
whereas $2$-layer networks saturate close to the origin. 
None of these behaviors matches the benchmark, 
showed in dashed lines, in this limit. 
We note that the $20$ realizations provide a relatively 
similar amount of deviations around a central value. 
This is in contrast to the $D-$state wavefunction of the $1$sc architecture, panel (e) of Fig.~\ref{fig:wf_var_wf}, which has a large variation around
the origin, in spite of providing the best overall energies. 

Unlike in Ref.~\cite{Keeble2020}, we display the wavefunctions of Fig.~\ref{fig:wf_var_wf} in a linear and denser grid of points 
than the one used to compute energy quadratures. This allows us to investigate whether the network is able to learn
efficiently not only the properties at the mesh points or the origin, but also the continuity of the wavefunction 
across a set of points at finite momentum. 
Indeed, we observe a step-like behavior in the two-layer $D-$state wavefunctions (panels (f) and (h) of Fig.~\ref{fig:wf_var_wf}). 
These steps occur precisely in the vicinity of each quadrature mesh point and they are 
more easily seen at high momenta where mesh points are relatively
spaced apart. Clearly, two-layer networks have enough flexibility to generate 
horizontal steps around these points. In other words, the networks learn the properties
of the wavefunction only locally around the meshpoints and
interpolate in a stepwise fashion between them. We take this as another, different 
sign of out-of-sample uncertainty.
We stress that the steps do not occur in  simpler, less flexible one-layer models. 
We speculate that the lack of flexibility of such models forces them to behave 
in a more continuous fashion. 
Further,  
these effects are not detected in integrated quantities, since the quadrature meshpoint 
values are well reproduced by
all the networks. 
In other words, in our set-up,
the regions between mesh points do not contribute to the loss functions.
Therefore, the network can push the variance to these regions with no
observable effect. We also note that all wave functions go immediately to zero for values of $q$ larger than the ones shown in Fig. \ref{fig:wf_var_wf}.

\begin{figure*}[t]
    \includegraphics[width=\linewidth]{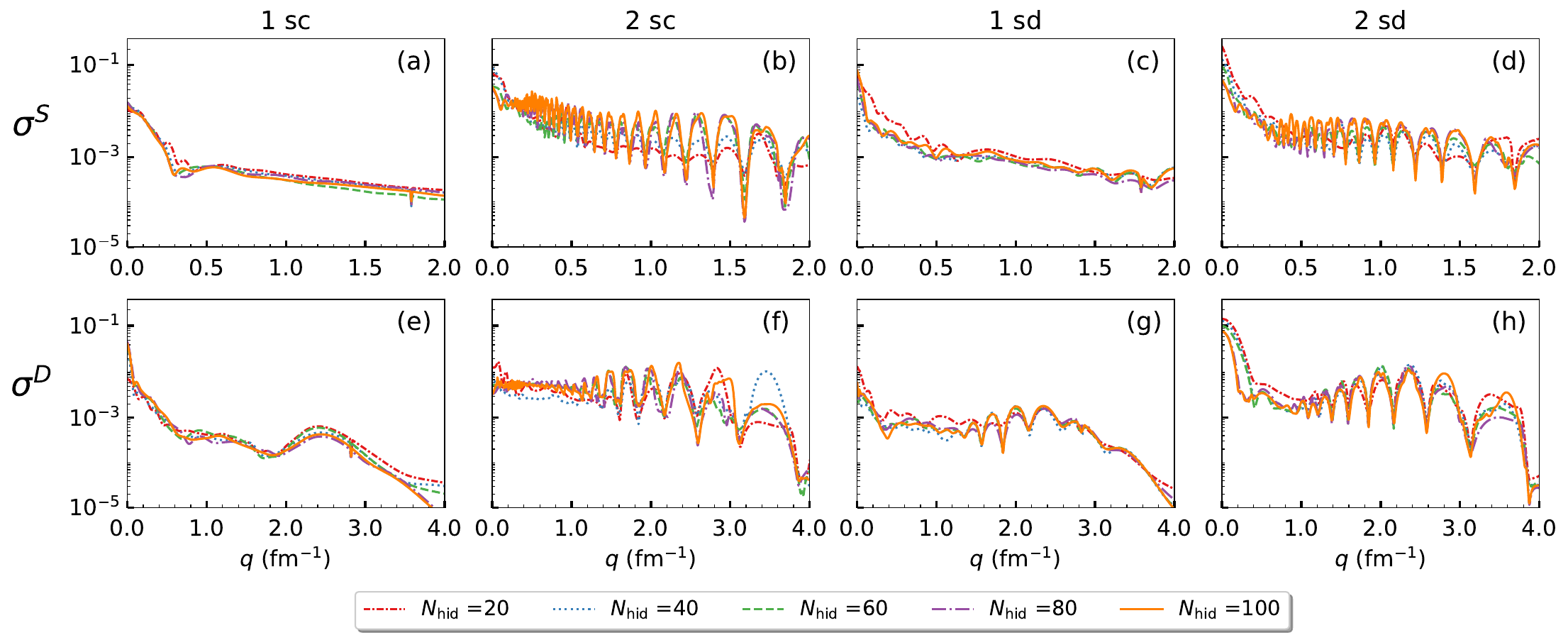}
    \caption{Top panels: standard deviation associated to $20$ instances of the $S-$state wavefunction as a function 
    of momentum, for each of the 4 architectures considered here. Different line styles correspond to different $N_\mathrm{hid}$ values. 
        Bottom panels: the same for the $D-$state wavefunction. Note the difference in scales of the momentum in both rows.
    }
    \label{fig:wf_var_stdev}
\end{figure*}

Figure~\ref{fig:wf_var_stdev} provides further insight into the structure 
of the wavefunction variance in our models. The variance here is
estimated as the standard deviation associated to the $20$ model initializations
of our networks. The different line styles in the figure correspond to
different values of $N_\text{hid}$. Top (bottom) panels show
$S$ ($D$) state variances. 
It is quite clear from these figures that the overall variance at 
finite momentum is relatively independent of the network width. 
Moreover, most network models have maximum variances towards the origin, which is in line with Fig.~\ref{fig:wf_var_wf}. 
$1-$layer networks have relatively flat declines with momentum, and saturate 
at values of the order of $\sigma \approx 10^{-3}$. 

In contrast, $2-$layer networks
have pronounced oscillatory values. The oscillation 
positions have a one-to-one correspondence with
the plateaus in Fig.~\ref{fig:wf_var_wf}. Variance minima occur at the quadrature
mesh points, and maxima happen in between. 
The oscillation minima (e.g. the smallest uncertainties) have variances 
which are similar to the $1-$layer case.
This is a clear indication that the network has learned locally 
the wavefunctions of the system at the quadrature mesh grids, 
but is overfitting the values in between which are not
penalised by the cost function. 
We discuss potential mitigation strategies to this problem
in the following sections. 

Before discussing solutions to this problem, we want to quantify
its relevance. To this end, we generate wavefunction models 
in a new uniform $10^3$-point mesh in the interval $[0,5]$ fm$^{-1}$.
We use this new linear mesh to compute, in quadrature, the overlaps of the
network models to the benchmark wavefunction as well as the total energy.
To summarise our findings, we show in Table~\ref{tab:overfitting_quantification} the
results of this exercise for 2 different models. On the one hand, 
the $1$sd model with $N_{\mathrm{hid}}=100$ 
(top row) shows almost no signs of overfitting. 
In contrast, the middle row shows results for the $2$sd 
architecture with $N_{\mathrm{hid}}=100$, which has significant overfitting. 
The bottom row quotes  the values obtained with a benchmark
wavefunction, obtained in the very same uniform mesh. 

\begin{table*}[t]
    \centering
    \begin{tabular}{c c c c} 
     \hline\hline
     & $F^S$ & $F^D$ & $E$ (MeV) \\ [0.5ex] 
     \hline
     $1$sd & $0.99997\pm 0.00001$ & $0.9999\pm 0.0001$ & $-2.216\pm 0.003$ \\ 
     $2$sd & $0.9997\pm 0.0001$ & $0.99\pm 0.01$  & $-2.1\pm 0.1$ \\
     Benchmark & 1 & 1 & -2.22486\\ 
     \hline\hline
    \end{tabular}
\caption{Fidelity and energies of the $1$sd and $2$sd architectures with $N_\text{hid}=100$ hidden neurons in a uniform mesh, where the overfitting is captured. 
Each value is the mean over the set of $20$ initializations, and the errors are the associated standard deviations.}
\label{tab:overfitting_quantification}
\end{table*}

First, we discuss the fidelities reported in the first and second columns of
Table~\ref{tab:overfitting_quantification}. 
These are computed in the uniform mesh, as opposed to the results presented
in Fig.~\ref{fig:energy_fidelity} for the original quadrature. 
For the $1$sd model, the fidelities
are practically the same, 
close to $1$ up to the fourth or fifth significant digit. 
This is due to the fact that the model has barely any overfitting.
The fidelities of the $2$sd model 
are, however, one and two orders of magnitude further away from one than those reported 
in Fig.~\ref{fig:energy_fidelity}. We take this as an indication of potential overfitting.

A second, stronger indication comes from the energies in the final column. 
These values should be compared to the benchmarks in the same uniform
mesh, reported in the bottom row. Two conclusions can be drawn here.
First, the energy values obtained in this mesh are significantly worse than
the exact ones.
Whereas for the $1$sd value
the energy lies within a fraction of a percent
of the benchmark, for the $2$sd model the model is only accurate 
up to $10 \%$ of the total energy. 
Second, the stochastic uncertainties are significantly larger than those 
reported in Fig.~\ref{fig:energy_fidelity}. 
For the $1$sd network, the values are relatively competitive, of 
the order of $1$ keV.  
The $2$sd model has stochastic uncertainties of $200$ keV, several
orders of magnitude higher. 

This substantial increase in stochastic uncertainty, as well as the important
change in energy central values, suggest that overfitting is particularly
problematic for $2-$layer networks. Among other things, 
this indicates that the results shown in Fig.~\ref{fig:energy_fidelity} 
are not fully representative of the true values of $2-$layer networks. 
The general behavior of these quantities as a function of $N_\mathrm{hid}$ is, 
however, expected to remain the same.

Having quantified the importance of overfitting, we now discuss its causes. 
The differences between the exact and predicted wavefunctions appear almost 
exclusively in two-layer models. These models have 
comparatively more parameters, which are presumably redundant.  
Two-layer networks can ``hide" this redundancy in regions with little 
or no contributions to the cost functions, either towards the origin or
in regions in-between fixed quadrature mesh points.
We stress that having too many parameters does not restrict the ANN outputs, but it may
increase the difficulty of the task assigned to the optimiser. 
The easy path out in this case is to overfit the mesh points. 

Moreover, as one increases the number of parameters,
the network may find multiple ways of fitting the data
set evaluated on the same fixed $N_q$ mesh points. 
In other words, larger networks can represent the same wavefunctions in many different 
ways, and as the number of parameters grow, so do the number of ways in which a network 
can represent the wavefunctions. 
This is usually identified as a bias-variance trade-off.

We now speculate on how to mitigate the overfitting effects on the wavefunctions. 
An obvious alternative is to try and increase the mesh density. One expects that in regions which are densely covered by the mesh, the networks may not have enough freedom to develop the artificial plateaus observed in Fig.~\ref{fig:wf_var_wf}. 
Just as in traditional variational models, improving the quality of the (momentum-space) basis by, say, increasing $N_q$, may also be beneficial in terms of the optimization. Alternatively, one can envisage a set-up in which meshes change from epoch to epoch. 
If such changes are entirely stochastic, one could rephrase them as traditional variational quantum Monte Carlo implementations, although these are not usually formulated with non-local interactions like the Entem-Machleidt potential. 
Alternatively, one could work with deterministic meshes but change the total number of meshpoints or the meshpoint positions at each epoch. In the most favourable scenario, one would expect to find a network that does not memorise specific data mesh points, but
rather learns the wavefunction structure. 
In the end, any such strategy would try to enforce a network learning process
that is independent of the specific quadrature that is used to compute the
global cost energy function. 

\subsection{Hyperparameter exploration}{\label{subsection:further_exploration_of_hyperparameter_space}}
In the analysis performed throughout the previous subsections we explored the performance of four ANN architectures with different numbers of hidden neurons, $N_\mathrm{hid}$, with the rest of the hyperparameters kept fixed. Now, for the sake of completeness, we present some results for the ground state energy computed with different hyperparameters. We discuss here the $1$sd architecture and show a qualitative discussion for the other architectures in the Appendix. All the hyperparameter exploration has been performed for networks with $N_\text{hid}=60$.

In this exploration, we look at the effect of the activation function and the learning rate. We use one of the best-behaved optimizers in the literature (RMSprop), and let some of its parameters vary to explore additional systematics. Specifically, we modify the smoothing constant ($\alpha$) and the momentum ($\mu$) of RMSprop. 
To explore the systematic dependence of the hyperparameters, we change one specific hyperparameter at a time, keeping the values of all other hyperparameters fixed to those used in Section~\ref{sec:results}.
Specifically, our baseline uses a Sigmoid activation function, a learning rate of $0.01$ and $\alpha=0.9$ and $\mu=0.0$. 

\begin{table}[t]
\resizebox{\columnwidth}{!}{
\begin{tabular}{c c c c} 
 \hline\hline
 & \multicolumn{2}{c}{$1$sd}\\ [0.5ex] 
 & Hyperparam & $E$ (MeV) & $r$ \\
 \hline
 \multirow{3}{*}{Act. fun.} 
 & Sigmoid 
 & $-2.22473^{+(0.00006, 0.0007)}_{-(0.00006, 0.0006)}$ & $66.9\%$ \\
 & Softplus
 & $-2.2220^{+(0.0004, 004)}_{-(0.0004, 0.002)}$ & $12.2\%(*)$ \\
 & ReLU 
 & $-2.2223^{+(0.0003, 0.004)}_{-(0.0003, 0.002)}$ & $13.2\%(*)$ \\
 \hline
 \multirow{3}{*}{lr} 
 & $0.005$ 
 & $-2.22598^{+(0.00002, 0.0002)}_{-(0.00002, 0.0002)}$ & $90.7\%$ \\
 & $0.01$ 
 & $-2.22473^{+(0.00006, 0.0007)}_{-(0.00006, 0.0006)}$ & $66.9\%$ \\
 & $0.05$ 
 & $-2.2231^{+(0.0004, 0.003)}_{-(0.0004, 0.001)}$ & $5.3\%(*)$ \\
 \hline
 \multirow{3}{*}{$\alpha$} 
 & $0.7$ 
 & $-2.22503^{+(0.00006, 0.00 9)}_{-(0.00006, 0.0006)}$ & $59.6\%$ \\
 & $0.8$ 
 & $-2.22488^{+(0.00008, 0.0008)}_{-(0.00008, 0.0006)}$ & $65.6\%$ \\
 & $0.9$ 
 & $-2.22473^{+(0.00006, 0.0007)}_{-(0.00006, 0.0006)}$ & $66.9\%$ \\
 \hline
 \multirow{2}{*}{$\mu$} 
 & $0.0$ 
 & $-2.22473^{+(0.00006, 0.0007)}_{-(0.00006, 0.0006)}$ & $66.9\%$ \\
 & $0.9$ 
 & $-2.2253^{+(0.0003, 0.002)}_{-(0.0003, 0.0009)}$ & $28.5\%$ \\
 \hline\hline
\end{tabular}}
\caption{Energy ($E$) and convergence rate ($r$) for different activation functions (Act. fun.), learning rates (lr), smoothing constants ($\alpha$) and momentum ($\mu$). The errors are displayed in the format: ($\mathrm{\epsilon_{stoch}}$, $\mathrm{\epsilon_{osc}}$), where $\mathrm{\epsilon_{stoch}}$ and $\mathrm{\epsilon_{osc}}$ are the stochastic and oscillation errors, respectively. All these results correspond to the $1$sd architercture and $N_\text{hid}=60$.} 
\label{tab:hyperparams}
\end{table}

Table~\ref{tab:hyperparams} shows numerical results of both the energy and the convergence rate for different combinations of the aforementioned hyperparameters, all of them sharing the same network architecture, $1$sd. Notice that for each hyperparameter class there is a row which is repeated: this corresponds to the baseline hyperparameter configuration, which is also the same one used in the previous sections, and displayed here to facilitate the comparison. In a similar spirit to that in Section~\ref{sec:method}, we use $N_\text{tot}=150$ trained models to compute all the convergence rates presented in Table~\ref{tab:hyperparams}. Nevertheless, not all hyperparameter configurations achieve $N_\text{con}=20$ converged models with this $N_\text{tot}$. In Table~\ref{tab:hyperparams}, we include an asterisk (*) next to the convergence rate of the models which need more runs. For all such models, the convergence rate is computed using $N_\text{tot}=1100$.

Concerning the changes in the activation function shown at the top of the table, we observe that our choice, the Sigmoid, has the highest convergence rate of $r = 66.9 \%$ for this architecture. This is contrast to $r<15\%$ for both the ReLU and Softplus functions, indicating that models have a harder time converging for these functions. Not only this, but Sigmoids also yield the lowest (and hence best) energy value. In contrast, ReLU and Softplus come at considerably higher energies.  

Models with lower learning rates are inherently slower to train. Because we work with a fixed number of iterations, a change in the learning can thus have a large effect in the results. To attenuate such artifices, we train models with a learning rate that is half that of our benchmark, $lr=0.005$, for $5\times 10^5$ epochs instead of $2.5\times 10^5$ epochs. We also explore models with larger learning rates ($lr=0.05$) for the same number of epochs than our baseline. We observe that both the energy and the convergence rate improve as the learning rate decreases. The associated energy uncertainties also decrease, and we note that the best results for $lr=0.005$ are incompatible with the baseline of $lr=0.01$, which is half as cheap in epochs. 

Finally, we explore the hyperparameter space associated to the optimiser in the bottom rows of Table~\ref{tab:hyperparams}. Our results indicate that variations in the smoothing constant $\alpha$ and the momentum $\mu$ do not have a significant impact on the metrics studied here. Perhaps the most robust conclusion is that adding momentum to RMSprop reduces the convergence rate of our simulations. The interested reader can find a similar table for the $1$sc, $2$sc and $2$sd configurations in the Appendix.

\section{Conclusions and future outlook}
\label{sec:conclusions}

In this work, we use vANNs to compute the ground state properties of
the deuteron with a realistic Hamiltonian. To this end, we discretise the
problem in a fixed mesh on the relative momentum coordinate. We use standard
ML tools, based on PyTorch, to pre-train our models and subsequently minimise
the energy for a fixed number of epochs. 

High-quality solutions for the variational wavefunction were already obtained
in Ref.~\cite{Keeble2020} with a similar set-up. 
We extend this work here in two directions, aimed
at identifying fundamental limitations of the vANN approach. First, we look at 
different ANN architectures, increasing the number of layers and treating the
connection to the output states in a connected or disconnected fashion. 
Second, we identify a new source of uncertainty associated to the oscillations
around the final energy minimum. Third, by carefully analysing the wavefunction 
outputs, we identify conditions in which finite-momentum overfitting arises. 

All vANNs models provide excellent results for the energies and fidelities, when compared to benchmark wavefunctions. 
By looking at the rate of model convergence for different architectures, we find a first qualitative sign that two-layer networks have a harder time minimising the energy than their one-layer counterparts. 
In terms of model uncertainties, the post-evolution oscillation
errors dominate over the stochastic initialization errors, but they
remain relatively small (of the order of $6$ to $8$ keV)
across a wide range of network widths. When it comes to the structure
of the network, state-connected networks, with output nodes that are
connected to the internal hidden layer, provide marginally better 
results than state-disconnected architectures.

Overall, we find that two-layer networks provide worse results than 
one-layer models. This may have been expected since the input layer is based on 
a single, positive-defined degree of freedom (the magnitude of the relative
momentum, $q$).
Central values of the energy are less attractive, and 
the stochastic and post-evolution uncertainties are larger. We also identify a 
dependence on the number of hidden nodes, which is absent in the one-layer case. 
Representing the wavefunctions of these models at grid points that are not used 
in the minimisation process, we find anomalous horizontal steps between mesh
points for the two-layer models. These are clear signs of network overfitting.
It seems that, during the minimisation process, the network is able to push 
some of the redundant dependence of parameters into these regions, which do 
not contribute to the energy cost function. This is analogous to the observation of a large variance (in terms of wavefunction values) around the origin, where the spherical $q^2$ factor allows the network to change values arbitrarily around $q \approx 0$ 
without affecting the total energy.
Unlike the situation at the origin, however, a change of integration mesh
for the energy can easily detect the degradation of the model associated to
overfitting at finite momentum. 
We find that, in the new mesh, the fidelities with respect to benchmark wavefunctions become worse. The energy values in a different mesh are also substantially less attractive. The associated stochastic uncertainty increases, reaching up to $300$ keV in some cases. 

While unveiling the internal behavior of the ANN is hard, 
the comparison between different architectures certainly sheds some 
light on the fundamental limitations of these variational methods.
When it comes to presenting wavefunctions that depend on a single continuous
variable, our results indicate that one-layer networks provide a better
starting point than the more complex two-layer approaches. This effect may
be due to the fixed (if long) total number of epochs, but other observations indicate
that overfitting arises much more easily in deeper networks. 
We have explored the hyperparameter and activation function dependence of our results. We find that our baseline model, for Sigmoid functions and moderate learning rates, are almost optimal. The only significant dependence is on the learning rate, which unfortunately requires running simulations for a higher number of epochs.

Similarly, our results suggest that the network can tackle states with different
quantum numbers in a fully connected configuration. 
In training ANNs to represent continuous quantum systems, 
non-fixed grid methods may provide superior learning capabilities. Overall,
this experience provides useful ideas to build more sophisticated nuclear 
models and tackle more difficult problems, including those of a many-body nature. 

\begin{table*}[t]
\resizebox{\linewidth}{!}{\begin{tabular}
{c c c c c c c c} 

 \hline\hline
 & & \multicolumn{2}{c}{$1$sc} & \multicolumn{2}{c}{$2$sc} & \multicolumn{2}{c}{$2$sd} \\ [0.5ex] 
 & Hyperp. & $E$ (MeV) & $r$ & $E$ (MeV) & $r$ & $E$ (MeV) & $r$ \\
 \hline
 \multirow{3}{*}{A. F.} 
 & Sigmoid 
 & $-2.225796^{+(0.000007, 0.001)}_{-(0.000007, 0.0004)}$ & $100.0\%$ 
 & $-2.2248^{+(0.0002, 0.004)}_{-(0.0002, 0.001)}$ & $60.3\%$  
 & $-2.2231^{+(0.0003, 0.004)}_{-(0.0003, 0.001)}$ & $23.2\%$
 \\
 & Softplus 
 & $-2.22534^{+(0.00001, 0.002)}_{-(0.00001, 0.0009)}$ & $100.0\%$ 
 & $-2.22490^{+(0.00009, 0.002)}_{-(0.00009, 0.0009)}$ & $98.0\%$ 
 & $-2.22512^{+(0.00006, 0.003)}_{-(0.00006, 0.0009)}$ & $51.7\%$ 
 \\
 & ReLU 
 & $-2.2235^{+(0.0002, 0.006)}_{-(0.0002, 0.002)}$ & $45.0\%$ 
 & $-2.2247^{+(0.0002, 0.003)}_{-(0.0002, 0.001)}$ & $68.2\%$ 
 & $-2.2232^{+(0.0003, 0.004)}_{-(0.0003, 0.001)}$ & $15.2\%$ 
 \\
 \hline
 \multirow{3}{*}{lr} 
 & $0.005$ 
 & $-2.226305^{+(0.000009, 0.0002)}_{-(0.000009, 0.0001)}$ & $100.0\%$ 
 & $-2.2255^{+(0.0003, 0.002)}_{-(0.0003, 0.0006)}$ & $72.9\%$ 
 & $-2.2245^{+(0.0002, 0.0007)}_{-(0.0002, 0.0004)}$ & $49.0\%$
 \\
 & $0.01$ 
 & $-2.225796^{+(0.000007, 0.001)}_{-(0.000007, 0.0004)}$ & $100.0\%$ 
 & $-2.2248^{+(0.0002, 0.004)}_{-(0.0002, 0.001)}$ & $60.3\%$ 
 & $-2.2231^{+(0.0003, 0.004)}_{-(0.0003, 0.001)}$ & $23.2\%$
 \\
 & $0.05$ 
 & $-2.226111^{+(0.000004, 0.0005)}_{-(0.000004, 0.0002)}$ & $86.5\%$ 
 & $-2.2249^{+(0.0004, 0.01)}_{-(0.0004, 0.001)}$ & $13.2\%$ 
 & $-2.2215^{+(0.0002, 0.008)}_{-(0.0002, 0.002)}$ & $3.0\%(*)$
 \\
 \hline
 \multirow{3}{*}{$\alpha$} 
 & $0.7$ 
 & $-2.22567^{+(0.00002, 0.0002)}_{-(0.00002, 0.0001)}$ & $100.0\%$
 & $-2.2228^{+(0.0004, 0.003)}_{-(0.0004, 0.002)}$ & $55.0\%$ 
 & $-2.2229^{+(0.0002, 0.0009)}_{-(0.0002, 0.0007)}$ & $38.4\%$
 \\
 & $0.8$
 & $-2.225742^{+(0.000008, 0.0003)}_{-(0.000008, 0.0002)}$ & $100.0\%$ 
 & $-2.2239^{+(0.0004, 0.002)}_{-(0.0004, 0.001)}$ & $49.0\%$ 
 & $-2.2234^{+(0.0003, 0.002)}_{-(0.0003, 0.0008)}$ & $46.4\%$
 \\
 & $0.9$ 
 & $-2.225796^{+(0.000007, 0.001)}_{-(0.000007, 0.0004)}$ & $100.0\%$ 
 & $-2.2248^{+(0.0002, 0.004)}_{-(0.0002, 0.001)}$ & $60.3\%$ 
 & $-2.2231^{+(0.0003, 0.004)}_{-(0.0003, 0.001)}$ & $23.2\%$
 \\
 \hline
 \multirow{2}{*}{$\mu$} 
 & $0.0$ 
 & $-2.225796^{+(0.000007, 0.001)}_{-(0.000007, 0.0004)}$ & $100.0\%$ 
 & $-2.2248^{+(0.0002, 0.004)}_{-(0.0002, 0.001)}$ & $60.3\%$ 
 & $-2.2231^{+(0.0003, 0.004)}_{-(0.0003, 0.001)}$ & $23.2\%$
 \\
 & $0.9$ 
 & $-2.226614^{+(0.000007, 0.0002)}_{-(0.000007, 0.00007)}$ & $100.0\%$ 
 & & $0.0\%$ 
 & $-2.22612^{+(0.00008, 0.001)}_{-(0.00008, 0.0005)}$ & $3.2\%(*)$
 \\
 \hline\hline
\end{tabular}}
\caption{The same as Table~\ref{tab:hyperparams}, but for the ANN architectures $1$sc, $2$sc and $2$sd.}
\label{tab:hyperparams2}
\end{table*}

\begin{acknowledgements}
We thank Bruno Juli\'a-D\'iaz for fruitful discussions. 
This work is supported by STFC, through Grants Nos 
ST/L005743/1 and ST/P005314/1; 
by grant PID2020-118758GB-I00 funded by MCIN/AEI/10.13039/501100\newline 011033;
by the ``Ram\'on y Cajal" grant RYC2018-026072 funded by MCIN/AEI /10.13039/501100011033 and FSE “El FSE invierte en tu futuro”; and by
the ``Unit of Excellence Mar\'ia de Maeztu 2020-2023" award to the Institute of Cosmos Sciences,
funded by MCIN/AEI/1b0.13\newline 039/501100011033
\end{acknowledgements}

\appendix

\section{Hyperparameter dependence exploration}{\label{appendix:hyperparams}}
In section \ref{subsection:further_exploration_of_hyperparameter_space} we briefly commented on the various hyperparameters in our models, and in Table~\ref{tab:hyperparams} we showed the effects of these variations upon the energy and the convergence rate for the $1$sd architecture. Here, we provide a similar analysis for the architectures $1$sc, $2$sc and $2$sd.

Table~\ref{tab:hyperparams2} shows numerical results of both the energy and the convergence rate for different values of the hyperparameters. We explore this dependence by changing one parameter at a time and keeping the remaining terms fixed. Here we find the same issue with convergence rates discussed in Section~\ref{subsection:further_exploration_of_hyperparameter_space}: for some hyperparameter configurations, using $N_\text{tot}=150$ is insufficient to guarantee a minimum of $N_\text{con}=20$ converged models. For such models, we use $N_\text{tot}=1100$ trained models instead, and we include an asterisk (*) next to their convergence rate in Table~\ref{tab:hyperparams2}.
We extract two relevant pieces of information from this table.
First, we find that $1$sc is the most stable architecture. Both the convergence rates and the energy values of this network are robust against changes of the hyperparameters. 
Second, we also find that a high convergence rate does not always entail a low ground-state energy. This can be easily understood realising that, while the trend for convergence rates appears to be linked almost exclusively to the architecture, the energy values depend instead on a wider range of factors. This is certainly something to bear in mind in future vANN calculations, where ANN architectures may have to be explored in detail. 

We also find a less direct but significant dependence on the learning rate. Just as for the $1$sd architecture, we find that lowering the learning rate by a factor of $2$ (and doubling the number of epochs) seems to lead to better variational minima. These also show slightly smaller uncertainties and improved convergence rates. 

For the $1$sc and $2$sd architerctures, the numbers in Table~\ref{tab:hyperparams2} also suggest that adding momentum results in models with lower (and hence better) energies. 
We also notice that two-layer architectures have much worse convergence rates than their one-layer counterparts for $\mu=0.9$. This may be understood with similar arguments as those presented in the main text. The $2$sc and $2$sd networks present overfitting, which suggests that there is an excessive number of hidden neurons. Therefore, parameter updates in these models are prone to having a greater impact on the overall wavefunction, which makes training less predictable, and increasing the momentum value is yet another step in that same direction. 

Table~\ref{tab:hyperparams2} also confirms some of the results we discussed in the main body of this manuscript. We observe that two-layer architectures have lower convergence rates than one-layer networks  throughout the entire range of hyperparameter values explored here. 
The same thing occurs for disconnected architectures, which have lower rates and somewhat worse energies than their counterparts. We take this as an indication that the tendencies are robust and reflect features linked to the network architecture itself, and not the training process details. 


\section*{Data Availability Statement}
The code used to produce all the results in this paper, including the figures appearing in the text, is available at the GitHub repository \newline \href{https://github.com/javier-rozalen/deuteron}{https://github.com/javier-rozalen/deuteron}. There are instructions and figures to facilitate the understanding and usage of the code. 


\bibliography{biblio}

\end{document}